\documentclass[sigconf]{acmart}

\copyrightyear{2021}
\acmYear{2021}
\setcopyright{acmcopyright}\acmConference[AIES '21]{Proceedings of the 2021
AAAI/ACM Conference on AI, Ethics, and Society}{May 19--21, 2021}{Virtual Event,
USA}
\acmBooktitle{Proceedings of the 2021 AAAI/ACM Conference on AI, Ethics, and
Society (AIES '21), May 19--21, 2021, Virtual Event, USA}
\acmPrice{15.00}
\acmDOI{10.1145/3461702.3462612}
\acmISBN{978-1-4503-8473-5/21/05}
\newcommand{\pact}{PACT}

\begin{document}
\fancyhead{}

\title{Envisioning Communities: A Participatory Approach \\ Towards AI for Social Good}


\author{Elizabeth Bondi,$^*$\textsuperscript{\rm 1} Lily Xu,$^*$\textsuperscript{\rm 1} Diana Acosta-Navas,\textsuperscript{\rm 2} Jackson A. Killian\textsuperscript{\rm 1}}
\thanks{*Equal contribution. Order determined by coin flip.}
% \email{ebondi@g.harvard.edu}
% \author{Lily Xu}
\email{{ebondi, lily_xu, jkillian}@g.harvard.edu,  diana_acosta_navas@hks.harvard.edu}
% \author{Diana Acosta-Navas, Jackson A. Killian}
% \author{Jackson A. Killian}
% \email{jkillian@g.harvard.edu}
\affiliation{%
  \institution{\textsuperscript{\rm 1}John A. Paulson School of Engineering and Applied Sciences, Harvard University}
  \streetaddress{29 Oxford St}
  \city{\textsuperscript{\rm 2}Department of Philosophy}
%   \state{Massachusetts}
  \country{Harvard University}
%   \postcode{02138}
}

\renewcommand{\shortauthors}{Bondi and Xu, et al.}

\begin{abstract}
Research in artificial intelligence (AI) for social good presupposes some definition of social good, but potential definitions have been seldom suggested and never agreed upon. The normative question of what AI for social good research should be ``for'' is not thoughtfully elaborated, or is frequently addressed with a utilitarian outlook that prioritizes the needs of the majority over those who have been historically marginalized, brushing aside realities of injustice and inequity. 
We argue that AI for social good ought to be assessed by the communities that the AI system will impact, using as a guide the capabilities approach, a framework to measure the ability of different policies to improve human welfare equity. Furthermore, we lay out how AI research has the potential to catalyze social progress by expanding and equalizing capabilities. We show how the capabilities approach aligns with a participatory approach for the design and implementation of AI for social good research in a framework we introduce called \pact, in which community members affected should be brought in as partners and their input prioritized throughout the project. We conclude by providing an incomplete set of guiding questions for carrying out such participatory AI research in a way that elicits and respects a community's own definition of social good.
\end{abstract}




\maketitle

\section{Introduction}
Artificial intelligence (AI) for social good, hereafter AI4SG, has received growing attention across academia and industry. Countless research groups, workshops, initiatives, and industry efforts tout programs to advance computing and AI ``for social good.'' Work in domains from healthcare to conservation has been brought into this category \cite{shi2020artificial,wang2019ai,kwok2019ai}. We, the authors, ourselves have endeavored towards AI4SG work as computer science and philosophy researchers.

Despite the rapidly growing popularity of AI4SG, social good has a nebulous definition in the computing world and elsewhere \cite{green2019good}, making it unclear at times what work ought to be considered social good. For example, can a COVID-19 contact tracing app be considered to fall within this lauded category, even with privacy risks? Recent work has begun to dive into this question of defining AI4SG \cite{floridi2020design,madaio2020co,floridi2018ai4people}; we will explore these efforts in more depth in Section~\ref{sec:critique}. 

However, we point out that context is critical, so no single tractable set of rules can determine whether a project is ``for social good.'' Instead, whether a project may bring about social good must be determined by those who live within the context of the system itself; that is, the community that it will affect. This point echoes recent calls for decolonial and power-shifting approaches to AI that focus on elevating traditionally marginalized populations \cite{mohamed2020decolonial,kalluri2020don,lewis2020indigenous,mhlambi2020from,birhane2020towards,whittaker2019disability}. 

\begin{figure}
\centering
\includegraphics[width=.7\columnwidth]{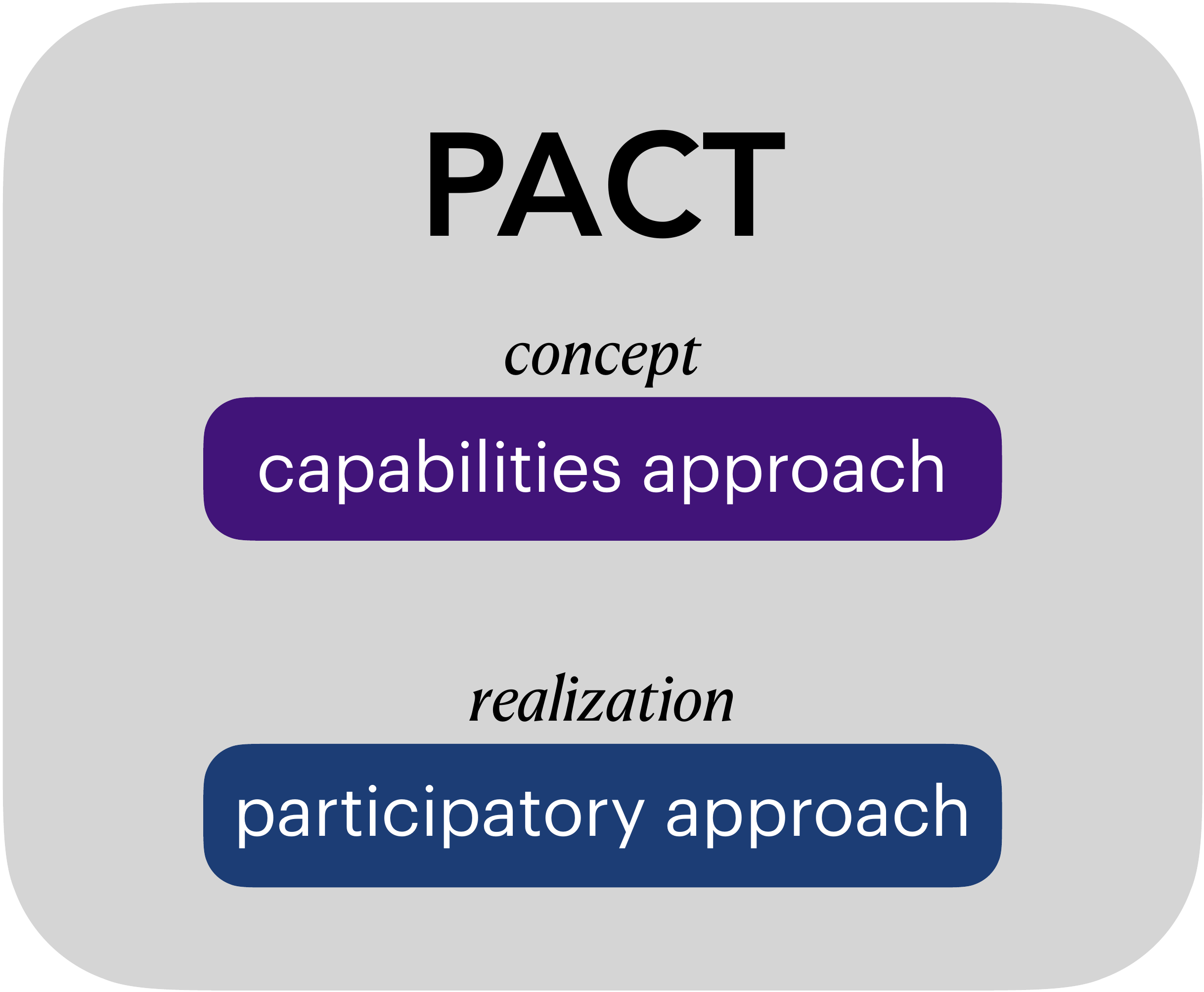}
\caption{The framework we propose, Participatory Approach to enable Capabilities in communiTies (PACT), melds the capabilities approach (see Figure~\ref{fig:capabilities}) with a participatory approach (see Figures \ref{fig:participatory} and \ref{fig:guiding-principles}) to center the needs of communities in AI research projects.}
\label{fig:pact}
\end{figure}

The community-centered, context-specific conception of social good that we propose raises its own questions, such as how to reconcile multiple viewpoints. We therefore address these concerns with an integrated framework, called the \textbf{P}articipatory \textbf{A}pproach to enable \textbf{C}apabilities in communi\textbf{T}ies, or \pact, that allows researchers to assess ``goodness" across different stakeholder groups and different projects. We illustrate \pact~in Figure~\ref{fig:pact}. As part of this framework, we first suggest ethical guidelines rooted in capability theory to guide such evaluations \cite{sen1999development,nussbaum2000feminism}. We reject a view that favors courses of action solely for their aggregate net benefits, regardless of how they are distributed and what resulting injustices arise. Such an additive accounting may easily err toward favoring the values and interests of majorities, excluding traditionally underrepresented community members from the design process altogether. 

Instead, we employ the \emph{capabilities approach}, designed to measure human development by focusing on the substantive liberties that individuals and communities enjoy, to lead the kind of lives they have reason to value \cite{sen1999development}. A capability-focused approach to social good is aimed at increasing opportunities for people to achieve combinations of ``functionings''---that is, combinations of things they may find valuable doing or being. While common assessment methods might overlook new or existing social inequalities if some measure of ``utility'' is increased high enough in aggregate, our approach would only define an endeavor as contributing to social good if it takes concrete steps toward empowering all members of the affected community to each enjoy the substantive liberties to function in the ways they have reason to value. 

We then propose to enact this conception of social good with a \emph{participatory approach} that involves community members in ``a process of investigating, understanding, reflecting upon, establishing, developing, and supporting mutual learning between multiple participants'' \cite{simonsen2012routledge}, in order for the community itself to define what those substantive liberties and functionings should be. In other words, communities define social good in their context.

Our contributions are therefore (i)~arguing that the capabilities approach is a worthy candidate for conceptualizing social good, especially in diverse-stakeholder settings (Section~\ref{sec:capabilities}), (ii)~ highlighting the role that AI can play in expanding and equalizing capabilities (Section~\ref{sec:AIandCapabilities}), (iii)~explaining how a participatory approach is best served to identify desired capabilities (Section~\ref{sec:participatory}), and (iv)~presenting and discussing our proposed guiding principles of a participatory approach (Section~\ref{sec:guiding-principles}). These contributions come together to form \pact.

\section{Growing Criticisms of AI for Social Good} \label{sec:critique}

As a technical field whose interactions on social-facing problems are young but monumental in impact, the field of AI has yet to fully develop a moral compass. On the whole, the subfield of AI for social good is not an exception. 

We highlight criticisms that have arisen against AI4SG research, which serve as a call to action to reform the field. Later, we argue that a participatory approach rooted in enabling capabilities will provide needed direction to the field by letting the affected communities---particularly those who are most vulnerable---be the guide. 

Recent years have seen calls for AI and computational researchers to more closely engage with the ethical implications  of  their  work. \citet{green2018data} implores researchers to view themselves and their work through a political lens, asking not just how the systems they build will impact society, but also how even the problems and methods they choose to explore (and \textit{not} explore) serve to normalize the types of research that ought to be done. \citet{latonero2019} offers a critical view of technosolutionism as it has recently manifested in AI4SG efforts emerging from industry, such as Intel's TrailGuard AI that detects poachers in camera trap images, which have the potential to individually identify a person. \citeauthor{latonero2019} argues that while companies may have good intentions, they often lack the ability to gain the expertise and local context required to tackle complex social issues. In a related spirit, \citet{blumenstock2018don} urges researchers not to forget ``the people behind the numbers'' when developing data-driven solutions, especially in development contexts. \citet{de2018machine} focus on a specific subset of AI4SG dubbed machine learning for development (ML4D) and similarly express the importance of considering local context to ensure that researcher and stakeholder goals are aligned. 

Also manifesting recently are meta-critiques of AI4SG specifically, which contend that the subfield is vaguely defined, to troubling implications. \citet{moore2019ai} focuses specifically on how the choice of the word ``good'' can serve to distract from potentially negative consequences of the development of certain technologies, retorting that AI4SG should be re-branded as AI for ``not bad''.  
\citet{whatisErin2019} argues that AI4SG's imprecise definition hurts its ability to progress as a discipline, since the lack of clarity around what values are held or what progress is being made hinders the ability of the field to establish specific expertise.
\citet{green2019good} points out that AI4SG is sufficiently vague to encompass projects aimed at police reform as well as predictive policing, and therefore simply lacks meaning. \citeauthor{green2019good} also argues that AI4SG's orientation toward ``good'' biases researchers toward incremental technological improvements to existing systems and away from larger reformative efforts that could be better.

Others have gone further, calling for reforms in the field to say that ``good'' AI should seek to shift power to the traditionally disadvantaged. \citet{mohamed2020decolonial} put forth a decolonial view of AI that suggests that AI systems should be built specifically to dismantle traditional forms of colonial oppression. They provide examples of how AI can perpetuate colonialism in the digital age, such as through algorithmic exploitation via Mechanical Turk--style ``ghost work'' \cite{gray2019ghost} or through algorithmic dispossesion in which disadvantaged communities are designed \emph{for} without being allowed a seat at the design table. They also offer three ``tactics'' for moving toward decolonial AI, namely: (1)~a critical technical practice to analyze whether systems promote fairness, diversity, safety, and mechanisms of anticolonial resistance; (2)~reciprocal engagements that engender co-design between affected communities and researchers; and (3)~a change in attitude from benevolence to solidarity, that again necessitates active engagement with communities and grassroots organizations. In a similar spirit, \citet{kalluri2020don} calls for researchers to critically analyze the power structures of the systems they design for and consider pursuing projects that empower the people they are intended to help. For example, researchers may seek to empower those represented in the data that enables the system, rather than the decision-maker who is privileged with a wealth of data. This could be accomplished by designing systems for users to audit or demand recourse based on an AI system's decision \cite{kalluri2020don}. 

In tandem with these criticisms, there have been corresponding efforts to define AI4SG.
\citet{floridi2018ai4people} provide a report of an early initiative to develop guidelines in support of AI for good. Therein, they highlight risks and opportunities for such systems, outline core ethical principles to be considered, and offer several recommendations for how AI efforts can be given a ``firm foundation'' in support of social good. 
\citet{floridi2020design} later expand on this work by proposing a three-part account, which includes a definition, a set of guiding principles, and a set of ``essential factors for success.'' They define AI4SG as ``the design, development, and deployment of AI systems in ways that (i)~prevent, mitigate, or resolve problems adversely affecting human life and/or the well-being of the natural world, and/or (ii)~enable socially preferable and/or environmentally sustainable developments.'' Note that this disjunctive definition captures a broad spectrum of projects with widely diverging outcomes, leaving open still the possible critiques discussed above. 
However, their principles and essential factors contribute to establishing rough guidelines for projects in the field, as others have done \cite{tomavsev2020ai,madaio2020co} in lieu of seeking a definition.

\section{The Capabilities Approach}\label{sec:capabilities}

Clearly, AI for social good is in need of a stricter set of ethical criteria and stronger guidance. 
To understand what steps take us in the direction of a more equitable, just, and fair society, we must find the right conceptual tools and frameworks. Utilitarian ethics, a widely referenced framework, adopts the aggregation of utility (broadly understood) as the sole standard to determine the moral value of an action \cite{timmons2013moral,robichaud2005great}. 

According to classic utilitarianism, the right action to take in any given context is whichever action maximizes utility for society. This brand of utilitarianism seems particularly attractive in science-oriented circles, due to its claim that a moral decision can be made by simply maximizing an objective function. For example, in the social choice literature, \citet{bogomolnaia2005collective} argue for the use of utilitarianism as ``it is efficient, strategyproof and treats equally agents and outcomes.'' However, the apparent transparency of the utilitarian argument obscures its major shortcomings. First, like the problem of class imbalance in machine learning, a maximizing strategy will bias its objective towards the majority class. Hence, effects on minority and marginalized groups may be easily overlooked.  Second, which course of action can maximize utility for society overall cannot be determined on simple cost--benefit analyses. Any such procedure involves serious trade-offs, and a moral decision requires that we acknowledge tensions and navigate them responsibly. 

To take a recent example, the development of digital contact tracing apps in the wake of the COVID-19  pandemic represented a massive potential benefit for public health, yet posed a serious threat to privacy rights, and all the values that privacy protects---including freedom of thought and expression, personal autonomy, and democratic legitimacy. Which course of action will maximize utility? It is not just controversial. What verdict is reached will necessarily depend on what fundamental choices and liberties individuals value most, and which they are willing to forgo. Furthermore, for some groups the losses may be more significant than for others, and this fact requires due consideration.

A related, though distinct, standard might put aside the ambition of maximizing good and adopt some threshold condition instead. As long as it generates a net improvement over the \emph{status quo}, someone might say, an action can be said to promote social and public good. This strategy raises three additional problems: first, utility aggregates are insensitive to allocation and distribution; second, they are not sensitive to individual liberties and fundamental rights; and third, to the extent that they consider stakeholders' preferences, they do not take into account the idea that individuals' preferences depend on their circumstances and in better circumstances could have different preferences \cite{sen1999development}. 
% account for starting condition, analogous to prospect theory (Kahneman and Tversky)
%preference adaptation (the idea that individuals' preferences depend on their circumstances, and in better circumstances could have different preferences) and desire conditioning into account \cite{sen1999development}. 
Why should this matter? \emph{Because attempts to increase utility may be too easily signed off as progress while inflicting serious harm and contributing to the entrenchment of existing inequality.} 

For this reason, we believe that respect for liberty, fairness of distribution, and sensitivity to interpersonal variations in utility functions should serve as moral constraints, to channel utility increments towards more valuable social outcomes. Operating within these constraints suggests shifting the focal point from utility to a different standard. In the spirit of enabling capabilities, we suggest that such standard should be based on an understanding of the kinds of lives individuals have reason to value, how the distribution of resources in combination with environmental factors may create (or eliminate) opportunities for them to actualize these lives, and designing projects that create fertile conditions for these opportunities to be secured. 

This reorientation of the moral assessment framework is in line with a view of AI as a tool to shift power to individuals and groups in situations of disadvantage, as advocated by \citet{kalluri2020don}.  What this shift means, in operational terms, is not yet fully elucidated. 
We therefore hope to contribute to this endeavor by exploring conceptual tools that may guide developers in the process, providing intermediate landmarks. We consider ``capabilities'' to be suitable candidates for this task, as they constitute a step in the direction of empowering disadvantaged individuals to decide what, in their view, that shift should consist of.

\begin{figure}
\centering
\includegraphics[width=\columnwidth]{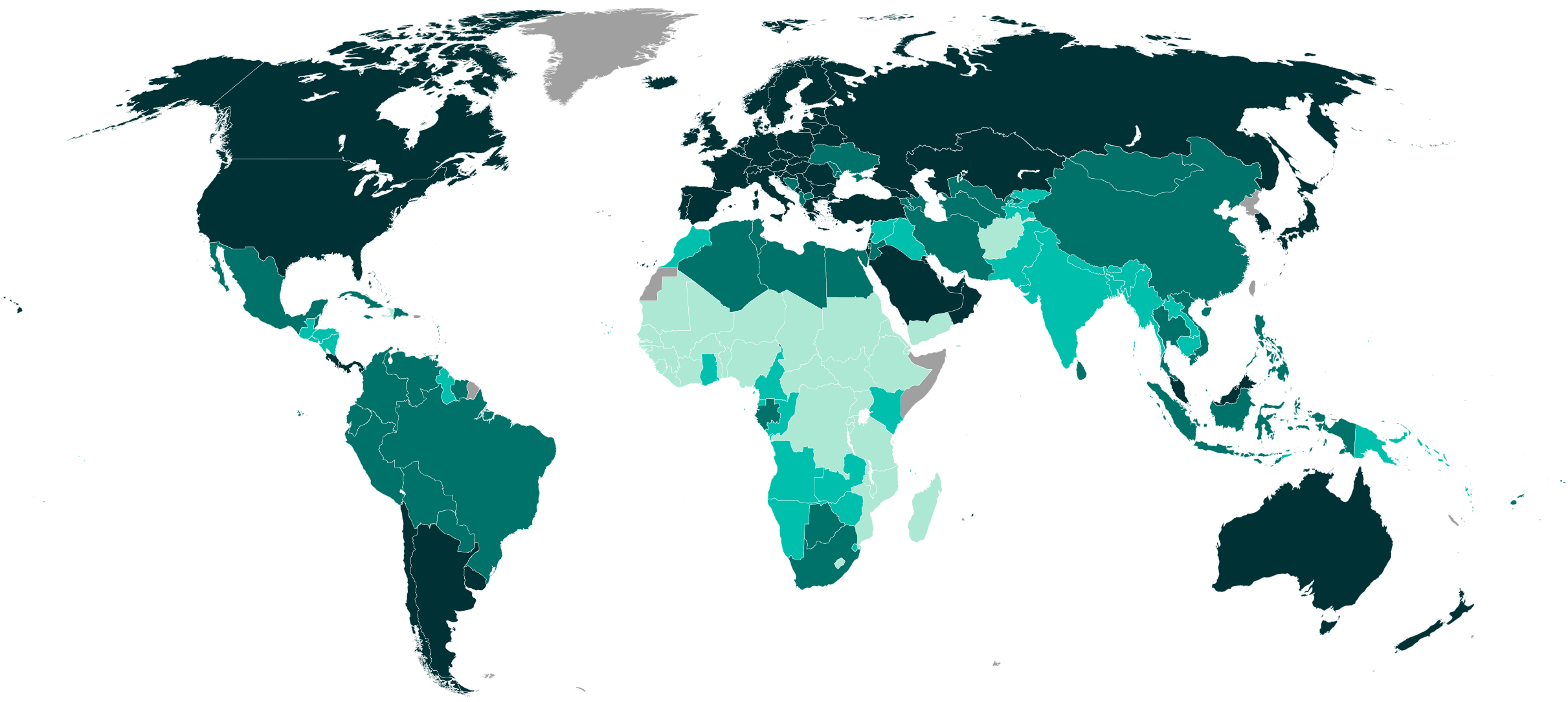}
\caption{The United Nations Human Development Index (HDI) was inspired by the capabilities approach, which serves as the foundation of the participatory approach that we propose. The above map from Wikimedia Commons shows the HDI rankings of countries around the world, where darker color indicates a higher index.}
\label{fig:hdi}
\end{figure}

\begin{figure*}
\centering
\includegraphics[width=.9\textwidth]{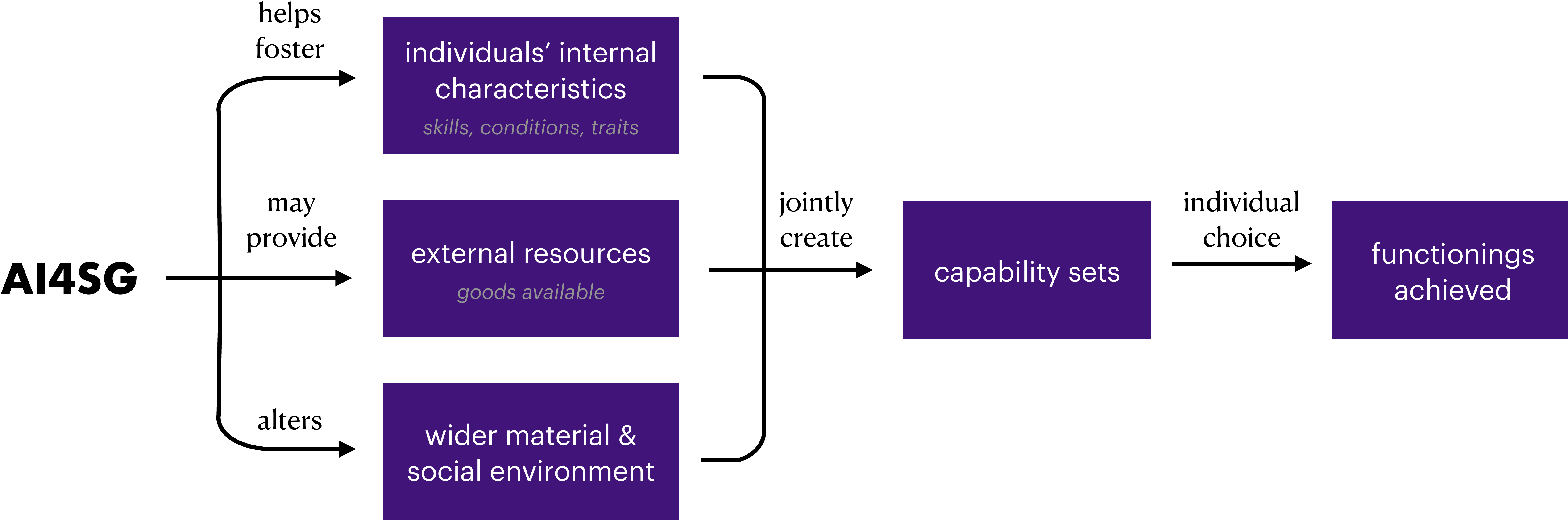}
\caption{AI for social good research projects can expand and equalize capabilities by fostering individuals' internal characteristics, providing external resources, or altering the wider material and social environment, thus creating capability sets to improve social welfare and fight injustice.}
\label{fig:capabilities}
\end{figure*}
The capabilities approach, which stands at the intersection of ethics and development economics, was originally proposed by Amartya Sen and Martha Nussbaum \cite{sen1999development,nussbaum2011creating}. The approach focuses on the notion that all human beings should have a set of substantive liberties that allow them to function in society in the ways they choose \cite{nussbaum2011creating}. These substantive liberties  include both freedom from external obstruction and the external conditions conducive to such functioning. The \textit{capability set} includes all of the foundational abilities a human being ought to have to flourish in society, such as, freedom of speech, thought and association; bodily health and integrity; and control over one's political and material environment; among others. The capabilities approach has inspired the creation of the United Nations Human Development Index (Figure~\ref{fig:hdi}), marking a shift away from utilitarian evaluations such as gross domestic product to people-centered policies that prioritize substantive capabilities \cite{ul2003birth}.

A capability-oriented AI does not rest content with increasing aggregate utility and respecting legal rights. It asks itself what it is doing, and how it interacts with existing social realities, to enhance or reduce the opportunities of the most vulnerable members of society to pursue the lives they have reason to value. It asks whether the social, political, and economic environment that it contributes to create deprives individuals of those capabilities or fosters the conditions in which they may enjoy equal substantive liberties to flourish. In this framework, a measure of social good is how much a given project contributes to bolster the enjoyment of substantive liberties, especially by members of marginalized groups. This kind of measure is respectful of individual liberties, sensitive to questions of distribution, and responsive to interpersonal variations in utility.

Before discussing the relation between AI and capabilities, we would like to dispel a potential objection to our framework. The capabilities approach has been criticized for not paying sufficient attention to groups, institutions, and social structures, thus rendering itself unable to account for power imbalances and dynamics \cite{hill2003development,koggel2003globalization}. This characteristic would make the approach unappealing for a field that seeks to address injustices in the distribution of power. However, the approach does indeed place substantial emphasis on conceptualizing the factors that affect (particularly those which may enhance) individuals' power. Its emphasis on substantial opportunities is itself a way of conceptualizing what individuals have the power to do or to be. These powers are determined in large part by the broader social environment, which includes group membership, inter-group relations, institutions, and social practices. Furthermore, the approach explicitly focuses on enhancing the power of the most vulnerable members of society. In fact, the creation and enhancement of capabilities constitute intermediate steps on the way towards shifting power, promoting the welfare and broadening the liberties of those who have been historically deprived. For this reason, it provides a fruitful framework to assess genuine moral progress and a promising tool for AI projects seeking to promote social good.

Light discussion of the capabilities approach has begun inside the AI community. \citet{moore2019ai} references the capabilities approach to call for greater accountability and individual control of private data. 
\citet{coeckelbergh2010health} proposes taking a capabilities approach to health care, particularly when using AI assistance to replace human care, but focuses on Nussbaum's proposed set of capabilities rather than eliciting desired capabilities from affected patients. We significantly build on these claims by discussing the potential of AI to enhance capabilities and argue that the capabilities approach goes hand-in-hand with a participatory approach. Two papers specifically ground their work in the capabilities approach; we highlight these as exemplars for future work in AI4SG. \citet{thinyane2019apprise} invoke the capabilities approach, specifically to empower the agency of marginalized groups, to motivate their development of a mobile app to identify victims of human trafficking in Thai fisheries.
\citet{kumar2020taking} conduct an empirical study of women's health in India through the lens of Nussbaum's central human capabilities.

\section{AI and Capabilities}\label{sec:AIandCapabilities}

The capabilities approach may serve AI researchers and developers to assess the potential impact of projects and interventions along two dimensions of social progress: capability expansion and distribution. This section argues that, when they interact productively with other social factors, AI projects can contribute to equalizing and enhancing capabilities---and that \emph{therein} lies their potential to bring about social good. For instance, an AI-based system Visual Question Answering System can enhance the capabilities of the visually impaired to access relevant information about their environment \cite{kim2019korean}.

When considering equalizing and enhancing capabilities, it is important to notice that whether an individual or group enjoys a set of capabilities is not solely a matter of having certain personal characteristics or being free from external obstruction. External conditions need also be conducive to enable individuals' choices among the set of alternatives that constitutes the capability set \cite{nussbaum2011creating}. This is why these liberties are described as \textit{substantial}, as opposed to formal or negative. Hence, capabilities, or substantive liberties, are  composed of (1)~individuals’ personal characteristics (including skills, conditions, and traits), (2)~external resources they have access to, and (3)~the configuration of their wider material and social environment \cite{johnstone2007technology}. 

Prior work at the intersection of technology and capabilities has addressed the potential of technology to empower users, by enhancing their capabilities and choice sets. \citet{johnstone2007technology} proposes the capabilities approach as a key framework for computer ethics, suggesting that technological objects may serve as tools or external resources that enhance individuals’ range of potential action and choice \cite{cohen2012configuring,kleine2013technologies}. This is an important role that AI-based technologies may play in enhancing capabilities: providing tools for geographic navigation, efficient communications, accurate health diagnoses, and climate forecasts. Technology may also foster the development of new skills, abilities, and traits that broaden an individual's choice sets. We diagram in Figure~\ref{fig:capabilities} the relationship between an AI4SG project on an individual's characteristics, resources, and environments, and thus the potential of AI to alter (both positively and negatively) capability sets. 

Technological objects may also become part of social structures, forming networks of interdependencies with people, groups, and other artifacts \cite{oosterlaken2015technology,winner1980artifacts}. When focusing on AI-based technologies, it is crucial to also acknowledge their potential to affect the social and material environment, which may render it more or less conducive to secure capability sets for individuals and communities. For example, AI-based predictive policing may increase the presence of law enforcement agents  in specific areas, which may in turn affect the capability sets of community members residing in those areas. If this presence increases security from, say, armed robbery, community members may enjoy some more liberties than they previously did. And yet, if this acts to reinforce the disparate impact of law enforcement on members of vulnerable communities, along with other inequalities, then it may not only diminish the substantive liberties of those individuals impacted, but also alter the way in which such liberties are distributed across the population.

Assessing this kind of trade-off ought to be a crucial step in evaluating the ability of a particular project to promote social good. This assessment must be aligned with the kinds of choices and opportunities community members have reasons to pursue \citep{sen2017collective}. A project should not be granted the moral standing of being ``for social good'' if it leads to localized utility increments, at the expense of reducing the ability of members of the larger community to choose the lives they value. Most importantly, this assessment must be made by community members themselves, as the following section argues.

\section{Community Participation} \label{sec:participatory}

As \citet{azra2021ai} contend, communities should ultimately be the ones to decide whether and how they would like to use AI. If the former condition is met and the community agrees that an AI solution may be relevant and useful, the latter requires the inclusive design of an AI system through a close and continued relationship between the AI researcher and those impacted.

This close partnership is particularly important as the gross effects of AI-based social interventions on communities’ capabilities are unlikely to be exclusively positive. Tradeoffs may be forced on designers and stakeholders; some are likely to be intergroup, and some, intragroup. For this reason, it is only consistent with the proposed approach that designers consult stakeholders from all groups on their preferred ways to navigate such tradeoffs. In other words, if \pact~is focused on creating capabilities, it must enable impacted individuals to have a say on what alternatives are opened (or closed) to them.

Moreover, AI projects that incorporate stakeholders' choices into their design process may contribute to create what \citet{wolff2007disadvantage} refer to as ``fertile functionings.'' That is, functionings that, when secured, are likely to secure other functionings. Fertile functionings include, though are not limited to, the ability to work and the ability to have social affiliations. These are the kinds of functionings that either enable other functionings (e.g. control over one's environment) or reduce the risk associated with them.  

Projects in AI that create propitious environments and enable individuals to make decisions over the capability sets they value in turn give those individuals the capability to function in a way that leads to the creation of other capabilities. If this kind of participatory space is offered to those who are the most vulnerable, AI could plausibly act against disadvantage, and contribute to shifting the distribution of power. In this way, the \pact~ framework is aligned with the principles of Design Justice in prioritizing the voices of affected communities and viewing positive change as a result of an accountable collaborative process \cite{2020Introduction}.

The \pact~framework is also committed to the notion that, when applying capabilities to AI, we must include all stakeholders, especially vulnerable members of society. But how do we accomplish this concretely, and how does this affect the different steps in an AI4SG research project? In the following section, we will first introduce suggested mechanisms of a participatory approach rooted in capabilities, and then discuss how these come into play in (1)~determining which capability sets to pursue at the beginning of a project, and (2)~evaluating the success of an AI4SG system in terms of its impact, particularly on groups' sets of substantive liberties. 

\begin{figure}
\centering
\includegraphics[width=\columnwidth]{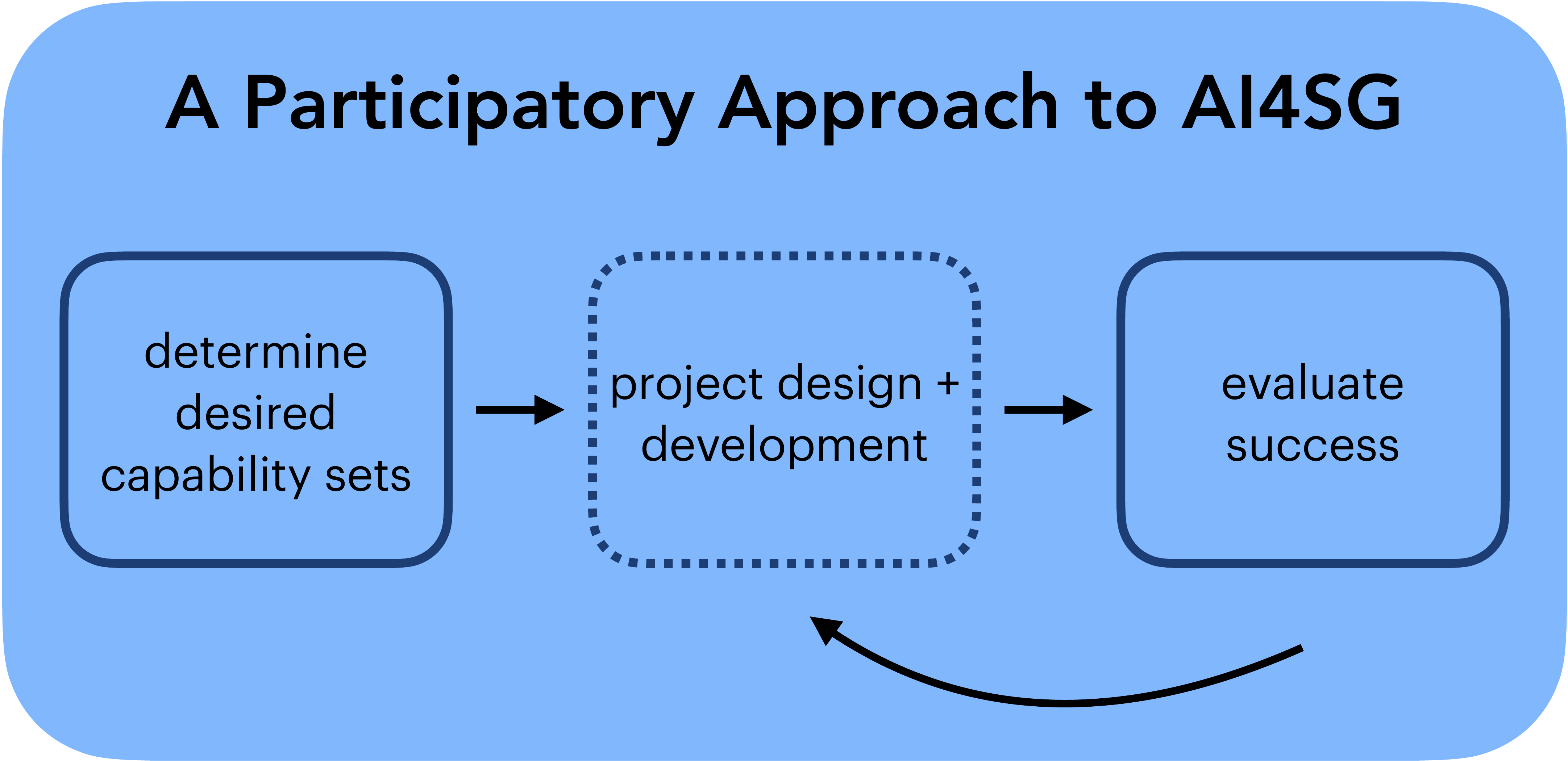}
\caption{Our proposed approach to AI4SG projects, where the key is that stakeholder participation is centered throughout. We do not explicitly comment on the design and development of the project other than calling for the inclusion of community participation and iterative evaluation of success, in terms of whether the project realized the desired capability sets.}
\label{fig:participatory}
\end{figure}

The approach we propose is diagrammed in Figure~\ref{fig:participatory}, which embeds community participation into each stage. By beginning first with defining capability sets, we resist the temptation to immediately apply established tools to address ingrained social issues, which would only restrict the possibilities for change \cite{lorde1984master}.
These participatory approaches constitute bottom-up approaches for embedding value into AI systems, by learning values from human interaction and engagement rather than being decided through the centralized and non-representative lens of AI researchers \cite{liao2019enabling}.

\section{Guiding Principles for a Participatory Approach}
\label{sec:guiding-principles}

\begin{figure*}
\centering
\includegraphics[width=\textwidth]{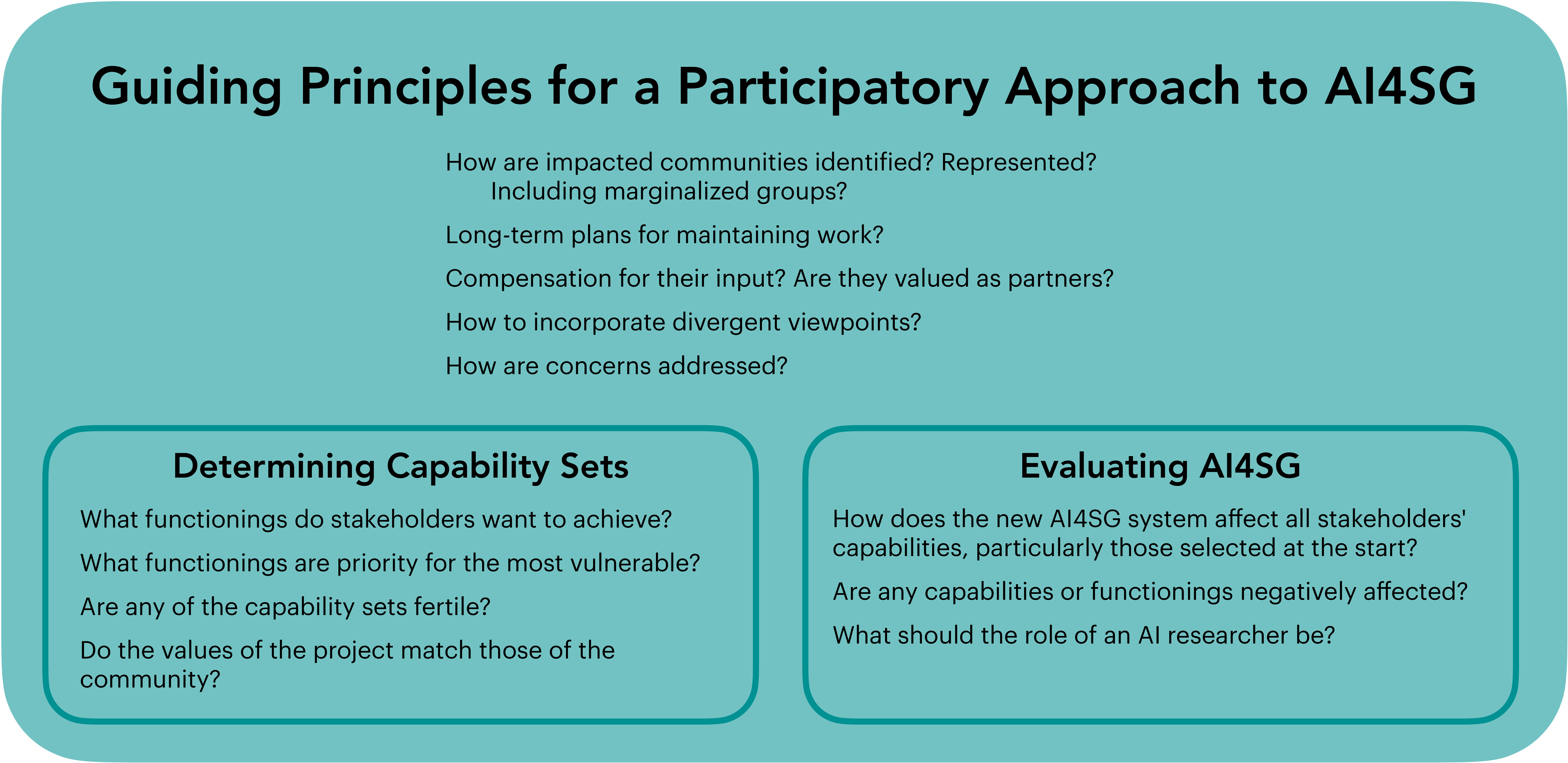}
\caption{Guiding principles for the participatory approach we propose. The key message is to center the community's desired capabilities throughout, and to ensure that the goals of the project are properly aligned to do so without negatively impacting other capabilities.}
\label{fig:guiding-principles}
\end{figure*}

To direct the participatory approach we propose, we provide the following set of guiding questions, outlined in Figure~\ref{fig:guiding-principles} and elaborated on below. These questions are specifically worded to avoid binary answers of yes or no. When vaguely stated requirements are put forth as list items to check off, there is a risk of offering a cheap seal of approval on projects that aren't sufficiently assessed in terms of their ethical and societal implications. Instead, we expect that in nearly all cases, none of these questions will be fully resolved. Thus, these questions are meant to serve as a constant navigation guide to help researchers regularly and critically reassess their intended work.

We begin our discussion on how a participatory approach to AI4SG would look with our first guiding question:

\smallskip
\noindent \textit{How are impacted communities identified and how are they represented in this process? Who represents historically marginalized groups?} 

\noindent This question is one of the most important and potentially difficult. As such, we encourage readers to research their domain of interest and seek partners as an important first step, but offer some advice from our experience. We often consider seeking partners at non-profits, community-based organizations, and/or \mbox{(non-)}governmental organizations (NGOs), and have even had some experience with non-profits or NGOs finding us. Finding these groups as an AI researcher may be done with the help of an institution's experts on partnerships (e.g., university or company officials focused on forming external partnerships), prior collaborators or acquaintances, discussions with peers within an institution in different departments (e.g., public health or ecology researchers), or ``cold emails.'' \citet{young2019toward} provide further guiding questions for finding these partners, particularly in their companion guide \cite{lassanaGuide}. In any case, some vetting and research is an important step. 

AI researchers may also co-organize workshops, such as a pair of ``AI vs.~tuberculosis'' workshops held by some of the authors in Mumbai, India which brought together domain experts, non-profits, state and local health officials, industry experts, and researchers, identified by Mumbai-based collaborators who specialize in building relationships across health-focused organizations in India. The varied backgrounds of the stakeholders at these workshops was conducive to quickly identifying areas of greatest need across the spectrum of tuberculosis care in India, many of which would not have been considered by AI researchers alone. Further, these group forums sparked conversations between stakeholders that often would not otherwise communicate, but that led to initial ideas for solutions. This highlights that such approaches are often most successful when AI researchers move out of their usual spheres and into the venues of the domains that their projects impact.

At the inception of a project, the designers and initial partners should together identify all relevant stakeholders that will be impacted by the proposed project, bringing them on as partners or members of a (possibly informal) advisory panel. Every effort should be made to include at least one member from each impacted group. If stakeholders were initially omitted and are later identified, they should then be added. Initial consultation with panel members and partners should be carried out in a way that facilitates open and candid discussion to ensure all voices are represented and accounted for during the project design phase \cite{young2019toward}. We additionally propose that during the lifetime of the project, the panel and partners should, if possible, be available for ongoing consultation during mutually agreed upon checkpoints to ensure the project continues to align with stakeholder values. Similar practices have been suggested by \citet{perrault2020ai}, who advocate for close, long-term collaborations with domain experts. Likewise, the Design Justice network pursues non-exploitative and community-led design processes, in which designers serve more as facilitators than experts. This framework is modeled on the idea of a horizontal relation in which affected communities are given the role of domain experts, given their lived experience and direct understanding of projects' impact \cite{2020Introduction}.

\smallskip
\noindent \textit{How are plans laid out for maintaining and sustaining this work in the long-term, and how would the partnership be ended?}

\noindent We believe partnership norms should be established at the beginning of the partnership, including ensuring that the community representatives and experiential experts have the power to end the project and partnership, in addition to the designers and other stakeholders. As noted by \citet{madaio2020co}, this should also necessitate an internal discussion amongst the research team to decide what, if any, criteria would be grounds for such a cessation, e.g., if it becomes clear that the problem at hand requires methods in which the research team does not have sufficient expertise. Discussions with the broader partners should also lay out the expected time scale and potential benefits and risks for all involved.

Once the relationship is underway, it should be maintained as discussed, with regular communication. When a project reaches the point where a potential deployment is possible, we advocate for an incremental deployment, such as the example of Germany's \textit{teststrecken} for self-driving cars, which sets constraints for autonomy, then expands these constraints once certain criteria are met \cite{floridi2020design}.

\smallskip
\noindent \textit{What kind of compensation are stakeholders receiving for their time and input? Does that compensation respect them as partners?}

\noindent Stakeholders must be respected as experiential or domain experts and be considered as partners. They must be compensated in some way, whether monetarily or otherwise. We advocate for compensation to avoid establishing an approach made under the guise of participatory AI that ends up being extractive \cite{pain2003reflections}. One notable drawback to a participatory approach is the potential for these community interactions to become exploitative if not done with intention, compensation, or long-term interactions \cite{sloane2020participation}. 
Collaborators who have significantly influenced the design or implementation of a project ought to be recognized as coauthors or otherwise acknowledged in resulting publications, presentations, media coverage, etc.

\smallskip
\noindent \textit{How can we understand and incorporate viewpoints from many groups of stakeholders?}

\noindent Surveys or voting may seem a natural choice. However, a simple voting mechanism is risky, as it may find that a majority of the community favors, for example, building a heavily polluting factory near a river, while the impacted community living at the proposed site would object that this factory would severely degrade their quality of life. These concerns from the marginalized group must be given due weight. This emphasis on the welfare of marginalized groups is based on the premise that the evaluation of human capabilities must consider \emph{individual capabilities} \cite{nussbaum2000feminism}. In short, all individual capabilities are valuable as ends in their own right; they should never be considered means to someone else's rights or welfare. We must, therefore, guard against depriving individuals of basic entitlements as a means to enhancing overall welfare. 

Hence, we endorse a deliberative approach, which aims to uncover ``overlapping consensus'' \cite{rawls1971theory}. This approach is based on the expectation that, in allowing diverse worldviews, value systems, and preference sets to engage in conversation, with appropriate moderation and technical tools, social groups may find a core set of decisions that all participants can reasonably agree with.
Deliberative approaches to democracy have been operationalized by programs such as vTaiwan, which uses digital technology to inform policy by building consensus through civic engagement and dialogue \cite{hsiao2018vtaiwan}. vTaiwan uses the consensus-oriented voting platform Pol.is which seeks to identify a set of common values upon which to shape legislation, rather than arbitrating between polarized sides \cite{tang2020inside}. Similarly, OPPi serves as a platform for consensus building through bottom-up crowd-sourcing \cite{oppi}. This tool is tailored for opinion sharing and seeking, oriented towards finding common ground among stakeholders.

An alternative to fully public deliberative processes are targeted efforts such as Diverse Voices panels \cite{young2019toward} which focus on including traditionally marginalized groups in existing policy-making processes. Specifically, they advocate for informal elicitation sessions with partners, asking questions such as, ``What do you do currently? What would support your work?'' \cite{katell2020toward}.
They also suggest considering whether tech should be used in the first place and highlight that a key challenge of participatory design is to determine a course of action if multiple participants disagree. We add that it remains a challenge to assemble these panels.

\citet{simonsen2012routledge} provide multiple strategies as well, particularly with the goal of coming to a common language and fostering communication between groups of experts from different backgrounds. They suggest strategies to invite discussion, such as games, acting out design proposals, storytelling, group brainstorms of what a perfect utopian solution would look like, participatory prototyping, and probing. In the case of probing, for example, one strategy was to provide participants with a cultural probe kit, consisting of items such as a diary and a camera, in order to understand people's reactions to their environments \cite{gaver1999design}.

Further, several fields in artificial intelligence have devoted a great deal of thought to learning and aggregating preferences: preference elicitation \cite{chajewska2000making}, which learns agent utility functions; computational social choice \cite{brandt2016handbook}, which deals with truthful agents; and mechanism design \cite{shoham2008multiagent}, which deals with strategic agents. As an example, \citet{kahng2019statistical} form what they call a ``virtual democracy'' for a food rescue program. They collect data about preferences on which food pantry should receive food, then use these preferences to create and aggregate the preferences of virtual voters. The goal is to make ethical decisions without needing to reach out to stakeholders each time a food distribution decision needs to be made. 

These fields have highlighted theoretical limitations in preference aggregation, such as Arrow's impossibility theorem \cite{arrow1950difficulty} which reveals limitations in the ability to aggregate preferences with as few as three voters, but these results do not necessarily inhibit us from designing good systems in practice \cite{sen1999possibility,maskin2014arrow}. Such areas provide rich directions for future research, particularly at the intersection of participatory methods and social science research. 

\smallskip
\noindent \textit{What specific concerns are raised during the deliberative process, and how are these addressed?}

Diverse Voices panels \cite{young2019toward} with experiential experts (both from non-profits and from within the community) may also be used to identify potential issues by asking questions such as ``What mistakes could be made by decision makers because of how this proposal is currently worded?'' and, ``What does the proposal not say that you wish it said?'' Other strategies we discussed for understanding multiple viewpoints may also have a role to play here when tailored towards sharing concerns. However, we also stress the importance of ongoing partnerships with impacted community members beyond just an initial session. By ensuring that the community has a voice throughout the project's lifetime, their values are always kept front-and-center. As others have argued, it is not sufficient to interview impacted communities once simply to ``check the participatory box'' \cite{sloane2020participation}.

\subsection{Determining Capability Sets}

Now, equipped with some guiding questions and concrete examples of a participatory approach, we will apply these principles directly to selecting capability sets at the outset of an AI4SG project. To identify what capabilities to work towards, some scholars such as \citet{nussbaum2000feminism} have proposed well-defined sets of capabilities and argued that society ought to focus on enabling those capabilities. However, we endorse the view that the selection of an optimal set of capabilities should be based on community consensus \cite{sen1999development}. We argue that an AI project can only be for social good if it is responsive to the values of the communities affected by the AI system.

\smallskip
\noindent \textit{What functionings do members of the various stakeholder groups wish they could achieve through the implementation of the project?}

In the fair machine learning literature, \citet{martin2020participatory} propose a participatory method called community-based system dynamics (CBSD) to bring stakeholders in to help formulate a machine learning problem and mitigate bias. Their method is designed to understand causal relationships, specifically feedback loops, particularly those in high-stakes environments such as health care or criminal justice, that may disadvantage marginalized and vulnerable groups. 
This process is intended to bring in relevant stakeholders and recognize that their lived experience makes these participants more qualified to recognize the effects of these interventions. Using visual diagrams designed by impacted community members, the CBSD method can help identify levers that may help enable or inhibit functionings, particularly for those who are most vulnerable. Similarly, \citet{simonsen2012routledge} suggest the use of mock-ups and prototypes to facilitate communication between experiential experts and developers. Other strategies discussed for understanding multiple viewpoints may also apply if tailored towards determining capabilities.

\smallskip
\noindent \textit{What functionings are the priority for those most vulnerable? Is there an overlap between their priorities and the goals of other stakeholders?}

We need to pay attention to those who are most vulnerable. The capabilities approach may be leveraged to fight inequality by thinking in terms of capability equalizing. To identify capabilities that are not yet available to marginalized members of a community, we must listen to their concerns and ensure those concerns are prioritized, for example via strategies proposed in our discussions on finding those impacted by AI systems and including multiple viewpoints. 

As an example of the consequences of failing to include those most vulnerable throughout the lifetime of an AI system, consider one of the initial steps of data collection. Women are often ignored in datasets and therefore their needs are underreported \cite{dignazio2020data}. %For example, voice-recognition software is trained on recordings of male voices, with commercial products 70\% less likely to understand women \cite{perez2019invisible}. But often these disparities are far more consequential. 
For example, crash-test dummies were designed to resemble the average male, and vehicles were evaluated to be safe based on these male dummies---leaving women 47\% more likely to be seriously injured in a crash \cite{perez2019invisible}. These imbalances are often also intersectional, as \citet{buolamwini2018gender} demonstrate by revealing stark racial and gender-based disparities in facial recognition algorithms. 

Beyond the inclusion of all groups in datasets, data must be properly stratified to expose disparities. During the COVID-19 pandemic in December 2020, the Bureau of Labor Statistics reported a loss of 140,000 net jobs. The stratified data reveal that all losses were women's: women lost 156,000 jobs while men gained 16,000, and unemployment was most severe for Black, Latinx, and Asian women \cite{ewing2021all}. 
Without accounting for the capabilities of all people affected by such systems, it is difficult to claim that these technologies were for social good.

On the other hand, prioritizing the needs of the most marginalized groups may at times offer an accelerated path towards achieving collective goals. Project Drawdown identified and ranked 100 practical solutions for stopping climate change \cite{hawken2017drawdown}. Number 6 on its list was the education of girls, recognizing that women with higher levels of education marry later and have fewer children, as well as manage agricultural plots with greater yield. Another solution advocates for securing land tenure of indigenous peoples, whose stewardship of the land fights deforestation, resource extraction, and monocropping.

\smallskip
\noindent \textit{Are any of these capability sets fertile, in the sense of securing other capabilities?}

To maximize the capabilities of various communities, we may wish to focus on capabilities that produce fertile functionings, as discussed in Section~\ref{sec:participatory}. 
Specifically, many functionings are necessary inputs to produce others; for example, achieving physical fitness from playing sports requires as input good health and nourishment \cite{clark2005sen}. Some of Nussbaum's 10 central capabilities---including bodily integrity (security against violence and freedom of mobility) and control over one's environment (right of political participation, to hold property, and dignified work)---may be viewed as fertile \cite{nussbaum2000feminism}. Reading, for instance, may secure the capability to work, associate with others, and have control over one's environment. AI has the potential to help achieve many of these capabilities. For example, accessible crowdwork, done thoughtfully, offers the opportunity for people with disabilities to find flexible work without the need for transit \cite{zyskowski2015accessible}.

\smallskip
\noindent \textit{How closely do the values of the project match those of the community as opposed to the designers? How does the focus of the project respond to their expressed needs and concerns? Does the project have the capacity to respond to those needs?}

\noindent Consider the scenario where a community finds it acceptable to hunt elephants, while the designers are trying to prevent poaching. There could be agreement on high-level values, such as public health, but disagreement on whether to prioritize specific interventions to promote public health. There could even be a complete lack of interest in the proposed AI system by the community. 

At an early stage of a project, AI researchers need to facilitate a consultation  method  to  understand  communities' values and choices. Note that this process could allow stark differences in priority between stakeholders to surface which prevent the project from starting so as not to over-invest in a project that will later be terminated because of difference in values. We may consider several of the strategies we discussed previously in this case, such as deliberative democratic processes, Diverse Voices panels, or computational social choice methods. It may subsequently be necessary to end the project if these strategies do not work.

\subsection{Evaluating AI for Social Good}

Once AI researchers and practitioners have a system tailored to these capabilities, we believe that communities should be the ones to judge the success of this new system. This stage of the process may pose additional challenges, given the difficulty of measuring capabilities \cite{johnstone2007technology}. Though we do not here endorse any measurement methodology, various attempts to operationalize the capabilities approach give us confidence that such methodologies are feasible and may be implemented in the course of evaluating AI projects \cite{anand2009development}.

First and foremost, we maintain that the evaluation of success should be done throughout the lifecycle of the AI4SG project (and beyond) as discussed above, especially via community feedback. However, we wish to emphasize that we as AI researchers need to keep capabilities in mind as we evaluate the success of AI4SG projects to avoid ``unintended consequences'' and a short-sighted focus on improved performance on metrics such as accuracy, precision, or recall.

\smallskip
\noindent \textit{How does the new AI4SG system affect all stakeholders' capabilities, particularly those selected at the start?}

\noindent This question is related to the literature on measuring capabilities \cite{johnstone2007technology}, and is therefore difficult to answer. We will aim to provide a few examples, which may not apply well to every AI4SG system, nor will it be an exhaustive list of techniques that could be valid. First, based on the idea of AI as a diagnostic to measure social problems \cite{abebe2020roles}, we may be able to (partially) probe this question using data. \citet{sweeney2013discrimination} show through ad data that advertisements related to arrest records were more likely to be displayed when searching for ``Black-sounding'' names, which may affect a candidate's employment capability, for example. \citet{obermeyer2019dissecting} analyze predictions, model inputs and parameters, and true health outcome data to show that the capability of health is violated for Black communities, as they are included in certain health programs less frequently than white communities due to a biased algorithm. 

There could additionally be feedback mechanisms when an intervention is deployed, whether via the AI system itself, or possibly by collaborating with local non-profits and NGOs. This may be especially useful in cases where these organizations are already engaged in tracking and improving key capabilities such as health outcomes, e.g., World Health Partners \cite{chavali2011world} or CARE international \cite{care2017annual}. 
Again, these examples will likely not apply to all cases and opens the door for further, interdisciplinary research. However, no matter what strategy is taken, it is imperative that we continue to center communities in all attempts to measure the effects on stakeholders' capabilities.
% In addition to traditional metrics of success in AI research, such as precision and recall, we also saw qualitative feedback from stakeholders to illustrate each of the capabilities they sought out to include.
%add Abebe et al - roles of computing for change, computing as diagnostic -- cite a few examples like obermeyer that show how AI can point out disparities -- however this is difficult
%move what role, should AI be used to determining capabilities section
%
% \smallskip
% \noindent \textit{How can we ensure that all stakeholders have access to each of the capabilities we selected at the start?}
%
% \noindent {\color{blue} maybe include reference here about measuring capabilities?} 
% This question will be extremely dependent on the particular domain, and is an area where collaboration with local non-profits and NGOs who operate in the community will be crucial. Such organizations can help measure the enhancement of the desired capabilities by adding observations or survey questions to data collection or monitoring procedures already in place.  

\smallskip
\noindent \textit{Are other valued capabilities or functionings negatively affected as a result of the project? Are stakeholders' values and priority rankings in line with such tradeoffs?}

\noindent We have a responsibility to \emph{actively} think about possible outcomes; it is neglect to dismiss negative possibilities as ``unintended consequences'' \cite{parvin2020unintended}. These participatory mechanisms should thus ensure that the perspective of the most vulnerable and most impacted stakeholders is given due consideration. We recognize that this can be especially challenging, as discussed in our first guiding question for identifying impacted communities. %if the population that would be negatively impacted was not already included in the development process. 
Therefore, we further suggest that the evaluation of an AI4SG project should employ consultation mechanisms that are open to all community members throughout the implementation process, such as the feedback mechanisms suggested previously.

\smallskip
\noindent\textit{What should my role be as an AI researcher? As a student?}

\noindent We believe that AI researchers at all levels should participate in this work. This work involves all of the above points, including learning about the domain to understand who stakeholders are,  discussing with the stakeholders, and evaluating performance. We also acknowledge that we AI researchers may not always be the best suited to lead participatory efforts, and so encourage interdisciplinary collaborations between computer science and other disciplines, such as social sciences. A strong example would be the Center for Analytical Approaches to Social Innovation (CAASI) at the University of Pittsburgh, which brings together interdisciplinary teams from policy, computing, and social work \cite{caasi}. However, AI researchers should not completely offload these important scenarios to social science or non-profit colleagues. It should be a team effort, which we believe will bring fruitful research in social science, computer science, and even more disciplines. 

AI researchers and students can also advocate for systematic change from within, which we discuss more in depth in the conclusion. Although student researchers are limited by constraints such as research opportunities and funding, they may establish a set of moral aspirations for their work and set aside time for people-centered activities, such as mentoring and community-building \cite{chan2020approaching}.

\section{Conclusion: Thoughts on AI for Social Good as a Field}

In this paper, we lay out a community-centered approach to defining AI for social good research that focuses on elevating the capabilities of those members who are most marginalized. This focus on capabilities, we argue, is best enacted through a participatory approach that includes those affected throughout the design, development and deployment process, and gives them ground to choose their desired capability sets as well as influence how they wish to see those capabilities realized. 

We recognize that the participatory approach we lay out requires a significant investment of time, energy, and resources beyond what is typical in AI or even much existing AI4SG research. \emph{We highlight this discrepancy to urge a reformation within the AI research community to reconsider existing incentives to encourage researchers to pursue more socially impactful work.} 

Institutions have the power to catalyze change by (1) establishing requirements for community engagement in research related to public-facing AI systems; and (2) increasing incentives for researchers to meaningfully engage impacted communities while simultaneously producing more impactful research \cite{black2020call}. 

While engaging in collaborative work with communities can give rise to some technical directions of independent interest to the AI community \cite{de2018machine}, such a shift to encourage community-focused work will in part require reconsidering evaluation criteria used when reviewing papers at top AI conferences. Greater value must be placed on papers with positive social outcomes, including those with potential for impact, if the work has not yet been deployed. Such new criteria are necessary since long-term, successful AI4SG partnerships often also lead to non-technical contributions, as well as situated programs which do not focus necessarily on generalizability \cite{perrault2020ai}. We encourage conferences to additionally consider rewarding socially beneficial work with analogies to Best Paper awards, such as the ``New Horizons'' award from the MD4SG 2020 Workshop, and institutions to recognize impactful work such as the Social Impact Award at the Berkeley School of Information \cite{berkeleyIschool}.

In the meantime, we suggest that researchers look to nontraditional or interdisciplinary venues for publishing their impactful community-focused work. These venues often gather researchers from a variety of disciplines outside computer science, opening the door for future collaborations. For example, researchers could consider COMPASS, IAAI, MD4SG/EAAMO \cite{abebe2018mechanism}, the LIMITS workshop \cite{nardi2018computing}, and special tracks at AAAI and IJCAI, among others. Venues such as the Computational Sustainability Doctoral Consortium and CRCS Rising Stars workshop bring students together from multiple disciplines to build relationships with each other. Researchers could also consider domain workshops and conferences, such as those in ecology or public health.

The incentive structure in AI research is often stacked against thoughtful deployment. Whereas a traditional experimental section may take as little as a week to prepare, a deployment in the field may take months or years, but is rarely afforded corresponding weight by reviewers and committee members. This extended timeline weighs most heavily on PhD students and untenured faculty who are evaluated on shorter timescales. 
We should thus reward both incremental and long-term deployment, freeing researchers from the pressure to rush to deployment before an approach is validated and ready. 

In addition to the need for bringing stakeholders into the design process of AI research, we must ensure that all communities are welcomed as AI researchers as well. Such an effort could counteract existing disparities and inequities within the field. For example, as is the case in other academic disciplines, systemic anti-Blackness is ingrained in the AI community, with racial discrepancies in physical resources such as access to a secure environment in which to focus, social resources such as access to project collaborators or referrals for internships, and measures such as the GRE or teacher evaluations \cite{guillory2020combating}. Further, as of 2018, only around 20\% of tenure-track computer science faculty were women \cite{roy2019engineering}. To combat these inequities, people across the academic ladder may actively work to change who they hire, with whom they collaborate, including collaborations with minority-serving institutions \cite{kuhlman2020no}, and how much time they spend on service activities to improve diversity efforts \cite{guillory2020combating}.

The above reformations could contribute greatly to making the use of participatory approaches the norm in AI4SG research, rather than the exception. 
\pact, we argue, is a meaningful new way to answer the question: ``what is AI for social good?'' We, as AI researchers dedicated to the advancement of social good, must make a \pact~with communities to find our way forward together.

\begin{acks}
We would like to thank Jamelle Watson-Daniels and Milind Tambe for valuable discussions. Thank you to all those we have collaborated with throughout our AI4SG journey for their partnership, wisdom, guidance, and insights. This work was supported by the Center for Research on Computation and Society (CRCS) at the Harvard John A. Paulson School of Engineering and Applied Sciences.
\end{acks}

\bibliographystyle{ACM-Reference-Format}
\bibliography{ref}

\end{document}